\documentclass[footinbib,aps,pra,reprint,superscriptaddress,nopacs]{revtex4-1}

\usepackage{graphicx}
\usepackage{amsmath,amsfonts,amssymb,bbm,bm}
\usepackage{verbatim}
\usepackage{braket}
\usepackage{mathtools}
\usepackage[absolute]{textpos}
\usepackage{ragged2e}

\usepackage{xcolor}

\begin{document}
\title{Adaptive bandwidth management for entanglement distribution in quantum networks}

\author{Navin B. Lingaraju}
\email{navin@purdue.edu}
\affiliation{School of Electrical and Computer Engineering and Purdue Quantum Science and Engineering Institute, Purdue University, West Lafayette, Indiana 47907, USA}
\author{Hsuan-Hao Lu}
\affiliation{School of Electrical and Computer Engineering and Purdue Quantum Science and Engineering Institute, Purdue University, West Lafayette, Indiana 47907, USA}
\author{Suparna Seshadri}
\affiliation{School of Electrical and Computer Engineering and Purdue Quantum Science and Engineering Institute, Purdue University, West Lafayette, Indiana 47907, USA}
\author{Daniel E. Leaird}
\affiliation{School of Electrical and Computer Engineering and Purdue Quantum Science and Engineering Institute, Purdue University, West Lafayette, Indiana 47907, USA}
\author{Andrew M. Weiner}
\affiliation{School of Electrical and Computer Engineering and Purdue Quantum Science and Engineering Institute, Purdue University, West Lafayette, Indiana 47907, USA}
\author{Joseph M. Lukens}
\affiliation{Quantum Information Science Group, Computational Sciences and Engineering Division, Oak Ridge National Laboratory, Oak Ridge, Tennessee 37831, USA}
\date{\today}

\begin{abstract}
Flexible-grid wavelength-division multiplexing is a powerful tool in lightwave communications for maximizing spectral efficiency. %, through agile provisioning of bandwidth tailored to the demands of individual communication channels.
In the emerging field of quantum networking,  %---tasked with establishing and distributing entanglement between many users---
the need for effective resource provisioning is particularly acute, given the generally lower power levels, higher sensitivity to loss, and inapplicability of digital error correction. %However, the power of adaptive bandwidth management has yet to be demonstrated in a multiuser quantum network.
In this Letter, we leverage flex-grid technology to demonstrate reconfigurable distribution of quantum entanglement in a four-user tabletop network. By adaptively partitioning bandwidth with a single wavelength-selective switch, we successfully equalize
%for adaptively partioning bandwidth
%, channels from a frequency-polarization hyperentangled biphoton source are sent to four users in a fully connected tabletop network. %, each with widely varying channel efficiencies.
%By adaptively partitioning bandwidth,
two-party coincidence rates that initially differ by over two orders of magnitude. %are equalized to within \fix{BLANK} across all users.
Our scalable approach introduces loss that is fixed with the number of users, offering a practical path for the establishment and management of quality-of-service guarantees in large quantum networks.
\end{abstract}

\maketitle

Quantum technology offers the promise of dramatic computational speed up~\cite{Ladd2010} and security~\cite{Gisin2007} beyond the capabilities of classical resources. Within this overall landscape, the development of quantum networks is critical for interconnecting quantum resources for applications such as blind quantum computing, quantum sensors, and distributed quantum computation~\cite{Wehner2018}. While the specific design of such networks remains an active area of research, %any solution will ultimately be tasked with entangling physically separated quantum nodes and facilitating secure communication between distant parties. Accordingly, 
to the greatest extent possible any solution should integrate seamlessly into the existing fiber-optic infrastructure, while also leveraging advanced techniques in modern lightwave communications. %The specific application of quantum key distribution (QKD) has spearheaded the expansion of quantum technology from lab to field, with many deployed QKD networks already demonstrated~\cite{Elliott2005, Peev2009, Chen2010, Sasaki2011, Wang2014, Zhang2018}. Nevertheless, the practical implementation of more general functionalities -- extending beyond QKD to, e.g., the distribution and verification of high-fidelity quantum entanglement for in principle arbitrary protocols -- represents an important need for quantum network development, particularly in the context of approaches that are agile, resource-efficient, and extendable to many users.

To this end, an efficient quantum networking approach gaining traction in recent years is based on entanglement distribution by a central provider~\cite{Lim2008,Brodsky2009,Herbauts2013,Ciurana2015,Aktas2016,Wengerowsky2018,Zhu2019}. In this paradigm, broadband polarization-entangled photons are carved into a series of spectral slices, which are then distributed to different users in the network. Since these photons are also entangled in the time-frequency degree of freedom, nonclassical polarization correlations are shared only between users who receive energy-matched channels. %Early work on this architecture examined the feasibility of this approach in term of entanglement quality across the biphoton bandwidth~\cite{Lim2008}, as well as in terms of advanced functionality such as users simultaneously maintaining multiple two-party links \cite{Brodsky2009}. Subsequent work introduced new functionality like active switching~\cite{Herbauts2013,Ciurana2015}, for example. 
Recently, Wengerowsky \emph{et al.}~\cite{Wengerowsky2018} demonstrated a fully and simultaneously connected QKD network by multiplexing multiple spectral slices to each user such that  %and doing so in a way that enabled a fully connected graph, i.e., a network where
polarization entanglement exists between every possible two-party link. This elegant demonstration relies only on passive components, namely, a hierarchical tree of dense wavelength-division multiplexing (DWDM) filters. %, thereby permitting distribution of polarization entanglement across the network. One limitation to the use of nested DWDM filters for entanglement distribution is that 
However, extending to significantly larger networks is a challenge, as the %For example, a network of only eight users would require over one hundred DWDM filters (
number of filters scales quadratically with the number of users $N$: a total of $2N^2-3N$ DWDMs are needed for full connectivity~\cite{Wengerowsky2018}. 
%104$, where $N = 8$ is the number of users).
%While no spectral slice passes through every filter, at least a few would have to pass through over ten DWDM filters –- effectively eliminating the advantage of the passive approach from the standpoint of loss.
Follow-on work~\cite{Joshi2020} made use of $50:50$ beam splitters to reduce the number of DWDMs in this architecture by having two users share each spectral slice. However, this comes at the expense of higher noise %as one can no longer recover the two-party coincidence-to-accidental ratio given
due to the intrinsic loss of 3~dB splitting. %Moreover, this improvement only halves what is a quadratic scaling in resources with the number of users. A further drawback of the nested DWDM approach is that the network is not dynamic in any meaningful sense. Adding and dropping users or altering how the spectrum is allocated requires physical rewiring of DWDM filters and beam splitters by the central provider. While the realization of a fully and simultaneously connected network architecture is appealing, it is an open question as to whether the advantage of using of only passive elements trumps scalability and dynamic reconfigurability –- features prized in classical lightwave communications.

\begin{figure}[tb!]
  \includegraphics[width=\columnwidth]{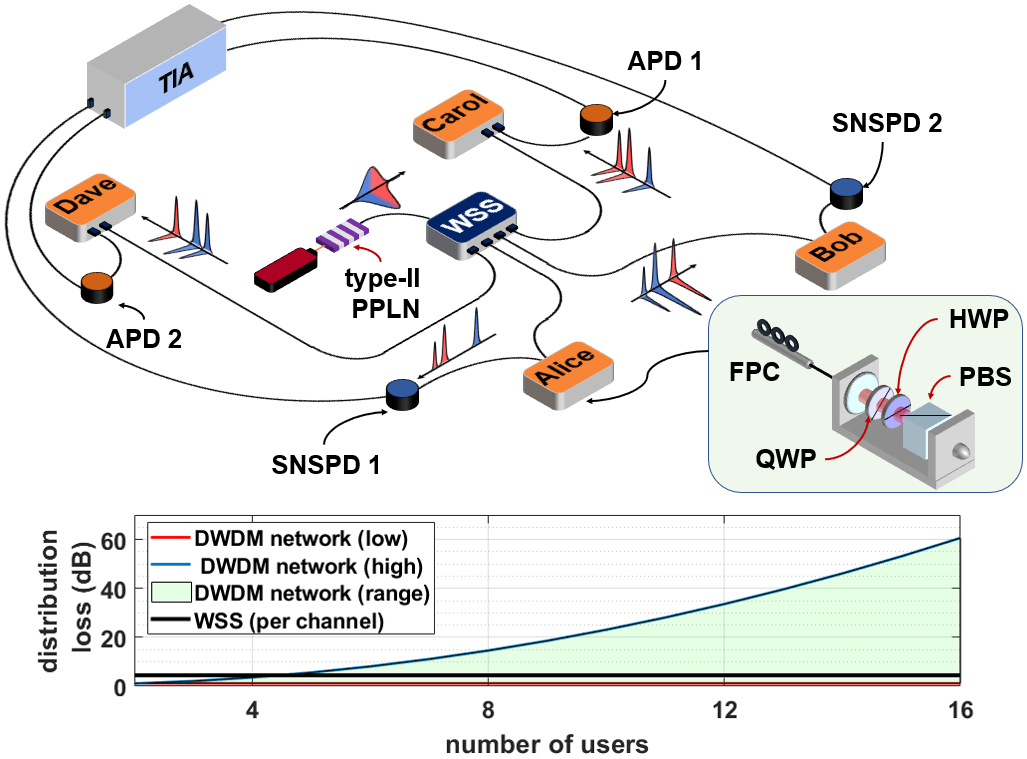}
  \caption{Network testbed for adaptive entanglement distribution. The inset shows the spread and scaling of channel losses for an alternative DWDM approach compared to the WSS. See text for details.}%CW laser, continuous-wave pump laser; PPLN, periodically-poled lithium niobate waveguide (HC Photonics); WSS, wavelength selective switch (Finisar); PC, polarization controller; $\frac{\lambda}{4}$, quarter-wave plate; $\frac{\lambda}{2}$, half-wave plate; PBS, polarizing beam splitter; SNSPD, superconducting nanowire single-photon detector (Quantum Opus); APD, single-photon avalanche diode (PicoQuant); TIA, time-interval analyzer.}
  \label{network}
\end{figure}

In this Letter, we propose and demonstrate a significantly improved approach to entanglement distribution. In lieu of passive optical elements, we use a wavelength-selective switch (WSS) to apportion the biphoton bandwidth between users on a network. This approach provides a clear advantage in terms of network scalability as the loss incurred during entanglement distribution is independent of the number of users, imparting no additional decoherence on the polarization-encoded quantum states due to the WSS's polarization-diversity design. Furthermore, the bandwidth allocation can be reconfigured dynamically with simple electronic control. Consequently, the central provider can not only enable a fully and simultaneously connected network, but any arbitrary subgraph. Lastly, the bandwidth of spectral slices routed to each user can be modified, thus making it possible to boost or throttle the entanglement rate for a particular two-party link without modifying the pair source, pump laser, or physical links on the network. Until there is broader deployment of quantum networks in general, what criteria or communications priorities should guide the distribution of entanglement will remain unclear. Yet our work highlights how the use of flexible and reconfigurable bandwidth provisioning can optimize network performance with regard to a preferred criterion or outcome.

%\section{Network Calibration and Characterization}
%\label{Network Arrangement}

\begin{figure*}[t!]
  \includegraphics[width=\textwidth]{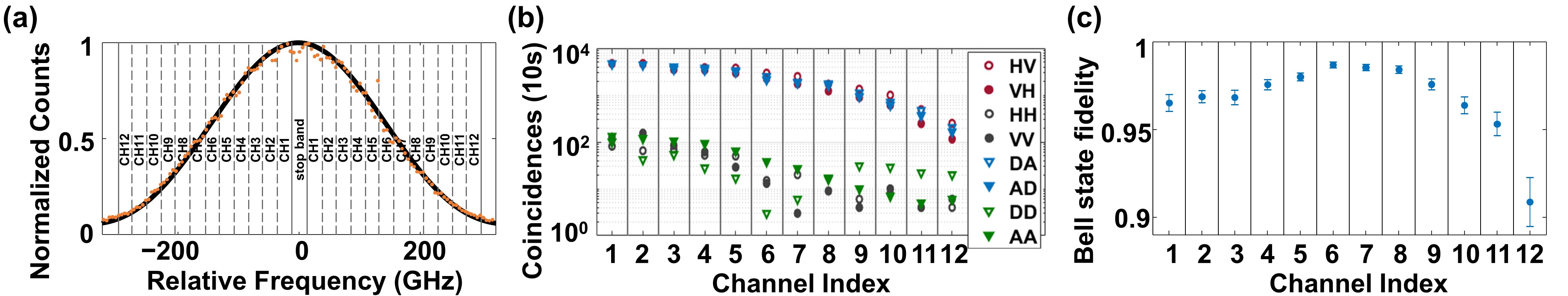}
  \caption{(a) Sinc-squared fit to the normalized singles rate as a function of detuning from the center of the biphoton spectrum. The locations of 12 pairs of energy-matched spectral slices are overlaid on the spectrum. (b) Data from polarization-correlation measurements in the rectilinear ($HV$) and diagonal ($DA$) bases, measured between Alice and Bob for all 12 channels. (c) Corresponding Bell state fidelities computed using Bayesian quantum state tomography.}
  \label{chacterization}
\end{figure*}

Our testbed is illustrated in Fig.~\ref{network}. Time-energy entangled photons are generated by spontaneous parametric down-conversion in a periodically poled lithium niobate ridge waveguide  (PPLN; HCPhotonics). The PPLN is engineered for type-II phase matching and is pumped by a $\sim$24 mW continuous-wave laser ($\lambda \approx 780$~nm)~\cite{Fujii2007}. In order to generate photon pairs also entangled in polarization, we match the pump wavelength and PPLN temperature to ensure spectrally degenerate down conversion~\cite{Zhu2013}. Temporal walk-off between horizontally and vertically polarized components of the biphoton %, which occurs both in the PPLN and the polarization-maintaining output coupler,
is compensated with a $90^{\circ}$ splice of polarization-maintaining (PM) fiber with proper length. %(to flip photons on the slow axis to the fast axis and vice versa). %followed by a specific length of PM fiber ($\approx$ 63 cm). found by the cutback method to optimize polarization entanglement).
As a result, any two energy-matched spectral slices $(\omega_\mathrm{signal} + \omega_\mathrm{idler} = \omega_\mathrm{pump})$ are polarization-entangled, ideally in the form $\ket{\Psi} \propto \ket{HV} + e^{i\phi}\ket{VH}$. %To confirm degenerate down conversion, the biphoton was sent to one port of a polarizing beam splitter with the state of polarization rotated from the horizontal-vertical ($HV$) basis to the diagonal-antidiagonal ($DA$) basis. Upon detuning the pump laser over a 2 nm range, we observe photon bunching at the point of degenerate down conversion with a visibility of $\approx 0.96$, suggesting a high degree of spatial, temporal, and spectral indistinguishablility between photons in a pair~\cite{Kuklewicz2004}.  
For entanglement distribution, the output of the PPLN is sent to a WSS, which can multiplex arbitrary spectral slices across the C- and L-bands to any one of four output ports. The only limitation on this programmability is the resolution of the WSS ($\sim$20 GHz), which sets a lower bound on the bandwidth of individual slices. %Photons are then carried from the output ports to four users by single-mode optical fiber. 
Subject to this bound, the WSS imparts equal losses for all connections up to the number of its output fibers.

On the other hand, the equivalent DWDM network presents widely varying loss across different channels as the number of users increases. While one channel need only travel through two DWDM filters, the worst-case spectral band must undergo, at the very least, $N^{2}-N$ reflections followed by 1 transmission~\cite{Wengerowsky2018}. As an example, Fig.~\ref{network} (inset) compares loss for fully connected networks assuming 4.5~dB loss for the WSS version (matching our specific device) and 0.25~dB (0.6~dB) for reflection (transmission) from each DWDM filter. Below four users, the DWDM approach is more efficient, but this rapidly changes as the network size increases, with a worst-case channel loss reaching 60~dB at the 16-user mark. We suspect it may be possible to design more balanced DWDM configurations that mitigate this wide loss spread, yet the fundamental quadratic scaling should still remain.

The users, identified as Alice, Bob, Charlie, and Dave (see Fig.~\ref{network}), are each equipped with a polarization analysis module that includes a fiber-based polarization controller (FPC), quarter-wave plate (QWP), half-wave plate (HWP), polarizer, and single-photon detector. The FPC compensates for rotation of the polarization state that occurs during transmission between the source and the QWP. While the manual FPC can map the state of polarization from the $HV$ basis of the source to the $HV$ basis of the polarization analysis module, there remains an undetermined phase between the $H$ and $V$ states.
%is no obvious manual way to compensate for phase accrued between two-photon basis states. %In other words, the biphoton state at the polarization analysis module is in the form $\ket{\Psi} \propto \ket{HV} + e^{i\phi}\ket{VH}$. Although polarization correlation measurements in the $HV$ basis are unaffected, one can only obtain high visibility in the DA basis for values $\phi\in \{0,\pi\}$.
We compensate for this unknown phase by orienting all QWPs at $45^{\circ}$, followed by additional rotation of HWP settings, for polarization correlation measurements in the diagonal-antidiagonal ($DA$) basis~\cite{Peters2005}. The HWP angle at which maximum contrast is obtained corresponds to the ``effective'' $DA$ basis; concretely, we set the angle such that the ideal quantum state is $\ket{\Psi^-} \propto \ket{HV}-\ket{VH}$. (We note that, with additional detectors for simultaneous $HV$ and $DA$ monitoring~\cite{Poppe2004} or automated feedback with classical reference pulses~\cite{Treiber2009}, it would be possible to compensate this directly with the FPC and remove the QWP.) Photons exiting the polarizer are routed to single-photon detectors for coincidence detection. %using single-mode optical fiber.
%An event timer tabulates single-photon events from all users, which we use to generate a histogram of two-photon delays. 

Of the four detectors used in this testbed, two are superconducting nanowire single-photon detectors (SNSPDs) while two are InGaAs avalanche photodiodes (APDs). The free-running SNSPDs used by Alice and Bob have quantum efficiencies $\sim$0.85. %and a deadtime of 30~ns.
The InGaAs APDs, which are allocated to Charlie and Dave, are gated with a 20 MHz clock ($10\%$ duty cycle) and have quantum efficiencies of $\sim$0.2 and $\sim$0.1, respectively. %, with deadtimes of 1000~ns.
%Figure~\ref{chacterization}(b) shows the singles rate at each detector
When the WSS is programmed to operate as a multiport ($1:4$) beam splitter for the whole biphoton bandwidth, the average singles count rates at each user are $2.6\times10^5$ s$^{-1}$ (Alice), $3.3\times10^5$ s$^{-1}$ (Bob), $5.5\times10^3$ s$^{-1}$ (Charlie), and $3.3\times10^3$ s$^{-1}$ (Dave), highlighting the vast disparity between the two classes of detectors. %It is clear that the singles rate is influenced not just by differences in detector efficiencies, but also by the use of gated detection with a CW pump in the case of APDs. 

Prior to apportioning the biphoton bandwidth for entanglement distribution over the network, we characterize the entanglement after the WSS using Alice's and Bob's polarization analysis modules. The biphoton spectrum is carved into 24 spectral slices, each of which is 24~GHz wide, and includes a central stopband [see Fig.~\ref{chacterization}(a)]. This channel width is chosen to match an integer multiple of the effective 4~GHz addressability of the WSS (owing to how pixels in the device's spatial light modulator are wired together) while still exceeding its $\sim$20~GHz spectral resolution. %The filter response of the WSS is programmed to correspond with the degeneracy point of the pump laser ($\frac{\omega_\mathrm{pump}}{2}$). In other words,
%Spectral slices that are symmetric with respect to the center of the biphoton spectrum are entangled in polarization.
We define a channel as a pair of such energy-matched spectral slices. %and there are such 12 channels across the the biphoton bandwidth. 
Fidelity with respect to the $\ket{\Psi^{-}}$ Bell state is determined by measurements in two sets of mutually unbiased bases [coincidences in $HV$ and $DA$ bases are shown in Fig.~\ref{chacterization}(b)]. Despite the tomographic incompleteness of this two-basis-pair set, our use of Bayesian mean estimation~\cite{Blume2010, Williams2017} nevertheless enables us to obtain meaningful state estimates, which due to the high correlations obtained also contain low uncertainty. Our measurement results using the Bayesian tomography workflow of Ref.~\cite{Lukens2020} are presented in Fig.~\ref{chacterization}(c). For channels 1--11, which span most of the biphoton bandwidth, we are able to measure fidelities higher than $95\%$, which illustrates both the quality of polarization rotation compensation and the stability of the WSS's polarization diversity scheme. %As has been noted in prior work~\cite{Lim2008,Herbauts2013}, there is a wavelength dependence to the unknown phase $\phi$ accrued between two-photon basis states, which ultimately limits the fidelity of polarization correlation measurements when using a single setting $\{\theta_{QWP}, \theta_{HWP}\}$ of the polarization analysis module for spectral slices across the entire biphoton bandwidth. 

%\section{Entanglement Distribution}
%\label{Entanglement Distribution}

\begin{figure*}[t!]
  \includegraphics[width=\textwidth]{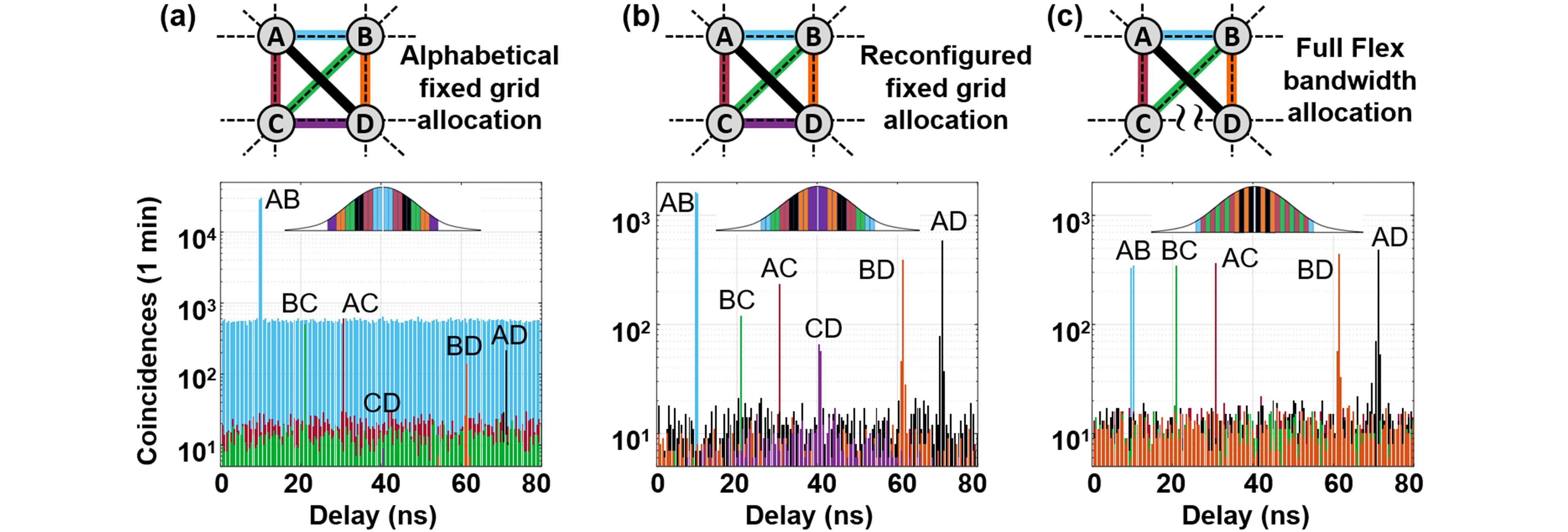}
  \caption{%\textit{(top left)} Illustration of the four-user QKD network with dashed lines denoting the possible ways in which nonclassical correlations in polarization state may be shared. %There are six unique user-to-user connections, or links, which have been numbered according to their pseudo-alphabetical ordering with each link assigned a unique color.
  (a) Coincidences for each two-party link based on a fixed 48~GHz grid, allocated alphabetically. The inset illustrates how the biphoton bandwidth is shared between all six links. (b) Coincidences for a fixed 48~GHz grid with channels allocated to best balance rates among all links. (c) Full-flex configuration with 24~GHz-wide spectral slices allocated freely between users to harmonize coincidence rates across a subgraph of the network. Link CD (APD--APD) is unable to be equalized and is dropped. %due to its  as this link would draw a disproportionate fraction of the biphoton bandwidth when equalizing the coincidence rates across all six two-party links
  }
  \label{distribution}
\end{figure*}

The four-user network illustrated in Fig.~\ref{network} comprises six possible two-party connections, called links, %Henceforth, a link is taken to mean a two-party user-to-user connection between whom nonclassical correlations in polarization state can be shared. %The top left map in Fig~\ref{distribution} is a representation of the physical layer showing all six links of the network.
each of which is assigned a unique color in Fig.~\ref{distribution}. The down-converted photons pass to the WSS for wavelength-multiplexed distribution to Alice, Bob, Charlie, and Dave, %Single-photon events at each detector are tabulated by the event timer to generate a histogram of the delay between the detection of two photons between a pair of users.
and we add suitable electronic offsets to the outputs of each detector to position the coincidence peaks for all six links in multiples of 10~ns apart. We use a %relatively large
coincidence window of 1024 ps, which exceeds the jitter of all detector pairs.
We first consider the case of wavelength-multiplexed entanglement distribution based on a fixed 48~GHz grid, corresponding to a total of 6 equal-width pairs from the 12 channels defined in Fig.~\ref{chacterization}(a). %except that now each link is defined by two contiguous channels.
 The biphoton bandwidth, counting from the center out, is allocated to the different links based on alphabetical ordering, i.e., Alice--Bob (AB), Alice--Charlie (AC), Alice--Dave (AD), Bob--Charlie (BC), Bob-Dave (BD), and Charlie--Dave (CD). Figure~\ref{distribution}(a) shows two-photon events recorded between all six links. %Coincidence windows are located at multiples of 10~ns. at delays of approximately 10 ns, 20 ns, 30 ns, 40 ns, 60ns, and 70ns.
Note the extremely low counts of link CD, due to the fact it combines the two least efficient channels and, under this distribution scenario, also receives the pair of 48~GHz spectral slices with the lowest flux. %From the perspective of equalizing two-party coincidence rates, this allocation is thus far from optimal. %for entanglement distribution based on a fixed spectral grid.
While far from optimal in terms of balancing two-party coincidence rates, this configuration might alternatively be interpreted as boosting service for a premium link, Alice--Bob in this case. Thus even in this example, we see the value of flexible bandwidth allocation in configuring the network for different needs, such as diverse quality-of-service targets.
%\subsection{Reconfigured fixed-grid allocation}
Nevertheless, to better balance coincidence rates across all six links, we next reallocate the biphoton bandwidth on the same fixed 48~GHz spectral grid. Now, brighter slices are routed to the less-efficient links. Figure~\ref{distribution}(b) shows histograms of two-photon events for all user-to-user connections. While the previous scenario saw a ratio of $\sim$4200 in the coincidence rates of links AB and CD, that imbalance is now only $\sim$26. 

%To be sure, this ability to manage coincidence rates derives not just from flexible allocation, but also from the fact that we are using close to the full biphoton spectrum. As Fig.~\ref{chacterization}(c) clearly shows, coincidence rates can differ by as much as a factor of 20 depending on which channels one is using. This use of the full down conversion spectrum marks a departure from previous works which limited consideration to spectral slices close enough to the center of the biphoton spectrum that there were minimal differences in the pair flux between different spectral slices. Our results show that in the case of links with disparate efficiencies there is utility in using all parts of the down converted spectrum and not just those in the flatband region.

%\subsection{Full-flex bandwidth provisioning}
Moving from fixed-grid distribution to a more flexible provisioning of bandwidth, we lastly utilize the 12 channels defined in Fig.~\ref{chacterization}(a), but do not place any limitations on how they are allocated between the different links. Aiming to equalize coincidence rates across the network, we provision the spectrum as shown in Fig.~\ref{distribution}(c); we no longer enable a fully connected network  because link CD would not even in the most favorable allocation (sending 7 channels with the highest pair flux) be able to reach coincidence rates comparable to the other users. Therefore, we program the WSS to harmonize the coincidence rate among a subgraph of five links (all links except for CD). From Fig. \ref{distribution}(c) we see that all coincidence rates are within a factor of 2, a significant improvement over the fixed-grid cases. 

Importantly, for the allocations of Fig.~\ref{distribution}(b-c), balancing coincidence rates relies not only on link bandwidth, but also use of the full biphoton spectrum [Fig.~\ref{chacterization}(a)] -- a departure from previous works which limited consideration to uniform-flux slices near the center of the spectrum~\cite{Wengerowsky2018,Joshi2020}. %As Fig.~\ref{chacterization}(b) shows, coincidence rates can differ by as much as a factor of 20 depending on the channel.
In our full-flex scenario in particular, interleaving channel allocations (e.g., for Links AC/BC and AD/BD) thereby allows us to equalize the mean flux for links of comparable detection efficiencies. While of course high flux across a broader bandwidth would always be preferred, our results show the value of flex-grid allocation under the constraints of limited entanglement resources in a network.

%\section{Discussion}
%\label{Discussion}
Moving forward, our approach should be readily extendable to many more users with only small modifications to the basic setup. While type-II phase-matching allows us to generate polarization-entangled photon pairs in a single-pass, single-crystal configuration, alternative sources based on type-0 or type-I nonlinear crystals (copolarized signal and idler) using Sagnac~\cite{Lim2008,Vergyris2017} or Mach--Zehnder ~\cite{Yoshizawa2004,Herbauts2013} arrangements would enable the generation of polarization-entangled photon pairs with essentially uniform flux across the entire C-band. Coupled with current commercial WSSs with 20 output ports, 4.8~THz bandwidth, and 6.25~GHz resolution~\cite{FinisarWSS}, a partitioning of 12.5~GHz slices would provide all $N(N-1)$ spectral channels necessary for a fully connected network of $N=20$ users. This stands in stark contrast to the nested DWDM approach~\cite{Wengerowsky2018}, for which increasing to 20 users would require a massive 740 DWDMs. Moreover, while a 20-user WSS could support full \emph{simultaneous} connectivity, its ability to reconfigure bandwidth means this is not required; the service provider can adjust bandwidth channels on demand to realize \emph{any} network subgraph, allocating unused bandwidth to increase entanglement rates in other channels. Such adaptivity makes the WSS approach superior in terms of latent resources compared to passive configurations.

In the present experiment, our quality-of-service metric focused on the coincidence rate between each pair of users. Although indicative of network performance, it does not directly reflect state fidelity -- crucial to quantum information protocols. In future work, through a combination of state tomography or entanglement witnesses, it should be possible to quantify the entanglement quality of each link in terms of ebits/s, which incorporates both detection rate and fidelity into a single metric, and then apportion bandwidth to equalize this target appropriately. Finally, while here we utilize frequency entanglement to establish bipartite correlations -- with polarization as the information carrier -- it is also possible to leverage the frequency degree of freedom directly for quantum information processing~\cite{Lukens2017,IEEEptl2019}. While typically considered in a single-polarization context, recent demonstrations of polarization-diversity phase modulation~\cite{Sandoval2019,Lingaraju2020} indicate the potential to exploit both frequency and polarization in parallel for carrying and processing quantum information for even more flexible quantum networks.

\medskip

\noindent\textbf{Funding.} U.S. Department of Energy, Office of Science, Office of Advanced Scientific Computing Research (Early Career Research Program); National Science Foundation (NSF) (1839191-ECCS, 1747426-DMR); Oak Ridge National Laboratory (Laboratory Directed Research and Development).

\medskip

\noindent\textbf{Acknowledgmenets.} Some preliminary results for this article were presented at CLEO 2020 as paper number FTh5D.2. We thank P. Imany, M. Alshowkan, B. Qi, and N.~A. Peters for discussions. We also thank N. Knight and O. E. Sandoval for helping with temporal walk-off compensation. A portion of this work was performed at Oak Ridge National Laboratory, operated by UT-Battelle for the U.S. Department of Energy under contract no. DE-AC05-00OR22725

\medskip

\noindent\textbf{Disclosures.} The authors declare no conflicts of interest.

\medskip

%\noindent See Supplement 1 for supporting content.
%\bigskip \noindent See \href{link}{Supplement 1} for supporting content.
\medskip


%\bibliography{references}

\begin{thebibliography}{26}%
\makeatletter
\providecommand \@ifxundefined [1]{%
 \@ifx{#1\undefined}
}%
\providecommand \@ifnum [1]{%
 \ifnum #1\expandafter \@firstoftwo
 \else \expandafter \@secondoftwo
 \fi
}%
\providecommand \@ifx [1]{%
 \ifx #1\expandafter \@firstoftwo
 \else \expandafter \@secondoftwo
 \fi
}%
\providecommand \natexlab [1]{#1}%
\providecommand \enquote  [1]{``#1''}%
\providecommand \bibnamefont  [1]{#1}%
\providecommand \bibfnamefont [1]{#1}%
\providecommand \citenamefont [1]{#1}%
\providecommand \href@noop [0]{\@secondoftwo}%
\providecommand \href [0]{\begingroup \@sanitize@url \@href}%
\providecommand \@href[1]{\@@startlink{#1}\@@href}%
\providecommand \@@href[1]{\endgroup#1\@@endlink}%
\providecommand \@sanitize@url [0]{\catcode `\\12\catcode `\$12\catcode
  `\&12\catcode `\#12\catcode `\^12\catcode `\_12\catcode `\%12\relax}%
\providecommand \@@startlink[1]{}%
\providecommand \@@endlink[0]{}%
\providecommand \url  [0]{\begingroup\@sanitize@url \@url }%
\providecommand \@url [1]{\endgroup\@href {#1}{\urlprefix }}%
\providecommand \urlprefix  [0]{URL }%
\providecommand \Eprint [0]{\href }%
\providecommand \doibase [0]{http://dx.doi.org/}%
\providecommand \selectlanguage [0]{\@gobble}%
\providecommand \bibinfo  [0]{\@secondoftwo}%
\providecommand \bibfield  [0]{\@secondoftwo}%
\providecommand \translation [1]{[#1]}%
\providecommand \BibitemOpen [0]{}%
\providecommand \bibitemStop [0]{}%
\providecommand \bibitemNoStop [0]{.\EOS\space}%
\providecommand \EOS [0]{\spacefactor3000\relax}%
\providecommand \BibitemShut  [1]{\csname bibitem#1\endcsname}%
\let\auto@bib@innerbib\@empty
%</preamble>
\bibitem [{\citenamefont {Ladd}\ \emph {et~al.}(2010)\citenamefont {Ladd},
  \citenamefont {Jelezko}, \citenamefont {Laflamme}, \citenamefont {Nakamura},
  \citenamefont {Monroe},\ and\ \citenamefont {O'Brien}}]{Ladd2010}%
  \BibitemOpen
  \bibfield  {author} {\bibinfo {author} {\bibfnamefont {T.~D.}\ \bibnamefont
  {Ladd}}, \bibinfo {author} {\bibfnamefont {F.}~\bibnamefont {Jelezko}},
  \bibinfo {author} {\bibfnamefont {R.}~\bibnamefont {Laflamme}}, \bibinfo
  {author} {\bibfnamefont {Y.}~\bibnamefont {Nakamura}}, \bibinfo {author}
  {\bibfnamefont {C.}~\bibnamefont {Monroe}}, \ and\ \bibinfo {author}
  {\bibfnamefont {J.~L.}\ \bibnamefont {O'Brien}},\ }\href {\doibase
  10.1038/nature08812} {\bibfield  {journal} {\bibinfo  {journal} {Nature}\
  }\textbf {\bibinfo {volume} {464}},\ \bibinfo {pages} {45} (\bibinfo {year}
  {2010})}\BibitemShut {NoStop}%
\bibitem [{\citenamefont {Gisin}\ and\ \citenamefont {Thew}(2007)}]{Gisin2007}%
  \BibitemOpen
  \bibfield  {author} {\bibinfo {author} {\bibfnamefont {N.}~\bibnamefont
  {Gisin}}\ and\ \bibinfo {author} {\bibfnamefont {R.}~\bibnamefont {Thew}},\
  }\href {\doibase 10.1038/nphoton.2007.22} {\bibfield  {journal} {\bibinfo
  {journal} {Nat. Photon.}\ }\textbf {\bibinfo {volume} {1}},\ \bibinfo {pages}
  {165} (\bibinfo {year} {2007})}\BibitemShut {NoStop}%
\bibitem [{\citenamefont {Wehner}\ \emph {et~al.}(2018)\citenamefont {Wehner},
  \citenamefont {Elkouss},\ and\ \citenamefont {Hanson}}]{Wehner2018}%
  \BibitemOpen
  \bibfield  {author} {\bibinfo {author} {\bibfnamefont {S.}~\bibnamefont
  {Wehner}}, \bibinfo {author} {\bibfnamefont {D.}~\bibnamefont {Elkouss}}, \
  and\ \bibinfo {author} {\bibfnamefont {R.}~\bibnamefont {Hanson}},\ }\href
  {\doibase 10.1126/science.aam9288} {\bibfield  {journal} {\bibinfo  {journal}
  {Science}\ }\textbf {\bibinfo {volume} {362}},\ \bibinfo {pages} {eaam9288}
  (\bibinfo {year} {2018})}  \BibitemShut {NoStop}%
\bibitem [{\citenamefont {Lim}\ \emph {et~al.}(2008)\citenamefont {Lim},
  \citenamefont {Yoshizawa}, \citenamefont {Tsuchida},\ and\ \citenamefont
  {Kikuchi}}]{Lim2008}%
  \BibitemOpen
  \bibfield  {author} {\bibinfo {author} {\bibfnamefont {H.~C.}\ \bibnamefont
  {Lim}}, \bibinfo {author} {\bibfnamefont {A.}~\bibnamefont {Yoshizawa}},
  \bibinfo {author} {\bibfnamefont {H.}~\bibnamefont {Tsuchida}}, \ and\
  \bibinfo {author} {\bibfnamefont {K.}~\bibnamefont {Kikuchi}},\ }\href
  {\doibase 10.1364/oe.16.022099} {\bibfield  {journal} {\bibinfo  {journal}
  {Opt. Express}\ }\textbf {\bibinfo {volume} {16}},\ \bibinfo {pages}
  {22099} (\bibinfo {year} {2008})}\BibitemShut {NoStop}%
\bibitem [{\citenamefont {Brodsky}\ and\ \citenamefont
  {Feuer}(2009)}]{Brodsky2009}%
  \BibitemOpen
  \bibfield  {author} {\bibinfo {author} {\bibfnamefont {M.}~\bibnamefont
  {Brodsky}}\ and\ \bibinfo {author} {\bibfnamefont {M.~D.}\ \bibnamefont
  {Feuer}},\ }\href@noop {} {\bibfield  {journal} {\bibinfo  {journal} {U.S.
  Patent Publication 2009/0180616 A1}\ } (\bibinfo {year} {2009})}\BibitemShut
  {NoStop}%
\bibitem [{\citenamefont {Herbauts}\ \emph {et~al.}(2013)\citenamefont
  {Herbauts}, \citenamefont {Blauensteiner}, \citenamefont {Poppe},
  \citenamefont {Jennewein},\ and\ \citenamefont {H{\"{u}}bel}}]{Herbauts2013}%
  \BibitemOpen
  \bibfield  {author} {\bibinfo {author} {\bibfnamefont {I.}~\bibnamefont
  {Herbauts}}, \bibinfo {author} {\bibfnamefont {B.}~\bibnamefont
  {Blauensteiner}}, \bibinfo {author} {\bibfnamefont {A.}~\bibnamefont
  {Poppe}}, \bibinfo {author} {\bibfnamefont {T.}~\bibnamefont {Jennewein}}, \
  and\ \bibinfo {author} {\bibfnamefont {H.}~\bibnamefont {H{\"{u}}bel}},\
  }\href {\doibase 10.1364/oe.21.029013} {\bibfield  {journal} {\bibinfo
  {journal} {Opt. Express}\ }\textbf {\bibinfo {volume} {21}},\ \bibinfo
  {pages} {29013} (\bibinfo {year} {2013})}\BibitemShut {NoStop}%
\bibitem [{\citenamefont {Ciurana}\ \emph {et~al.}(2015)\citenamefont
  {Ciurana}, \citenamefont {Martin}, \citenamefont {Martinez-Mateo},
  \citenamefont {Schrenk}, \citenamefont {Peev},\ and\ \citenamefont
  {Poppe}}]{Ciurana2015}%
  \BibitemOpen
  \bibfield  {author} {\bibinfo {author} {\bibfnamefont {A.}~\bibnamefont
  {Ciurana}}, \bibinfo {author} {\bibfnamefont {V.}~\bibnamefont {Martin}},
  \bibinfo {author} {\bibfnamefont {J.}~\bibnamefont {Martinez-Mateo}},
  \bibinfo {author} {\bibfnamefont {B.}~\bibnamefont {Schrenk}}, \bibinfo
  {author} {\bibfnamefont {M.}~\bibnamefont {Peev}}, \ and\ \bibinfo {author}
  {\bibfnamefont {A.}~\bibnamefont {Poppe}},\ }\href {\doibase
  10.1109/JSTQE.2014.2367241} {\bibfield  {journal} {\bibinfo  {journal} {IEEE
  J. Sel. Top. Quantum Electron.}\ }\textbf {\bibinfo {volume} {21}},\ \bibinfo
  {pages} {6400212} (\bibinfo {year} {2015})}\BibitemShut {NoStop}%
\bibitem [{\citenamefont {Aktas}\ \emph {et~al.}(2016)\citenamefont {Aktas},
  \citenamefont {Fedrici}, \citenamefont {Kaiser}, \citenamefont {Lunghi},
  \citenamefont {Labont{\'e}},\ and\ \citenamefont {Tanzilli}}]{Aktas2016}%
  \BibitemOpen
  \bibfield  {author} {\bibinfo {author} {\bibfnamefont {D.}~\bibnamefont
  {Aktas}}, \bibinfo {author} {\bibfnamefont {B.}~\bibnamefont {Fedrici}},
  \bibinfo {author} {\bibfnamefont {F.}~\bibnamefont {Kaiser}}, \bibinfo
  {author} {\bibfnamefont {T.}~\bibnamefont {Lunghi}}, \bibinfo {author}
  {\bibfnamefont {L.}~\bibnamefont {Labont{\'e}}}, \ and\ \bibinfo {author}
  {\bibfnamefont {S.}~\bibnamefont {Tanzilli}},\ }\href@noop {} {\bibfield
  {journal} {\bibinfo  {journal} {Laser Photonics Rev.}\ }\textbf
  {\bibinfo {volume} {10}},\ \bibinfo {pages} {451} (\bibinfo {year}
  {2016})}\BibitemShut {NoStop}%
\bibitem [{\citenamefont {Wengerowsky}\ \emph {et~al.}(2018)\citenamefont
  {Wengerowsky}, \citenamefont {Joshi}, \citenamefont {Steinlechner},
  \citenamefont {H{\"{u}}bel},\ and\ \citenamefont {Ursin}}]{Wengerowsky2018}%
  \BibitemOpen
  \bibfield  {author} {\bibinfo {author} {\bibfnamefont {S.}~\bibnamefont
  {Wengerowsky}}, \bibinfo {author} {\bibfnamefont {S.~K.}\ \bibnamefont
  {Joshi}}, \bibinfo {author} {\bibfnamefont {F.}~\bibnamefont {Steinlechner}},
  \bibinfo {author} {\bibfnamefont {H.}~\bibnamefont {H{\"{u}}bel}}, \ and\
  \bibinfo {author} {\bibfnamefont {R.}~\bibnamefont {Ursin}},\ }\href
  {\doibase 10.1038/s41586-018-0766-y} {\bibfield  {journal} {\bibinfo
  {journal} {Nature}\ }\textbf {\bibinfo {volume} {564}},\ \bibinfo {pages}
  {225} (\bibinfo {year} {2018})}\BibitemShut {NoStop}%
\bibitem [{\citenamefont {Zhu}\ \emph {et~al.}(2019)\citenamefont {Zhu},
  \citenamefont {Corbari}, \citenamefont {Gladyshev}, \citenamefont {Kazansky},
  \citenamefont {Lo},\ and\ \citenamefont {Qian}}]{Zhu2019}%
  \BibitemOpen
  \bibfield  {author} {\bibinfo {author} {\bibfnamefont {E.~Y.}\ \bibnamefont
  {Zhu}}, \bibinfo {author} {\bibfnamefont {C.}~\bibnamefont {Corbari}},
  \bibinfo {author} {\bibfnamefont {A.}~\bibnamefont {Gladyshev}}, \bibinfo
  {author} {\bibfnamefont {P.~G.}\ \bibnamefont {Kazansky}}, \bibinfo {author}
  {\bibfnamefont {H.-K.}\ \bibnamefont {Lo}}, \ and\ \bibinfo {author}
  {\bibfnamefont {L.}~\bibnamefont {Qian}},\ }\href {\doibase
  10.1364/josab.36.0000b1} {\bibfield  {journal} {\bibinfo  {journal} {J. Opt. Soc. Am. B}\ }\textbf {\bibinfo {volume} {36}},\
  \bibinfo {pages} {B1} (\bibinfo {year} {2019})}\BibitemShut {NoStop}%
\bibitem [{\citenamefont {Joshi}\ \emph {et~al.}(2020)\citenamefont {Joshi},
  \citenamefont {Aktas}, \citenamefont {Wengerowsky}, \citenamefont {Lon{\v
  c}ari{\'c}}, \citenamefont {Neumann}, \citenamefont {Liu}, \citenamefont
  {Scheidl}, \citenamefont {Lorenzo}, \citenamefont {Samec}, \citenamefont
  {Kling}, \citenamefont {Qiu}, \citenamefont {Razavi}, \citenamefont {Stip{\v
  c}evi{\'c}}, \citenamefont {Rarity},\ and\ \citenamefont
  {Ursin}}]{Joshi2020}%
  \BibitemOpen
  \bibfield  {author} {\bibinfo {author} {\bibfnamefont {S.~K.}\ \bibnamefont
  {Joshi}}, \bibinfo {author} {\bibfnamefont {D.}~\bibnamefont {Aktas}},
  \bibinfo {author} {\bibfnamefont {S.}~\bibnamefont {Wengerowsky}}, \bibinfo
  {author} {\bibfnamefont {M.}~\bibnamefont {Lon{\v c}ari{\'c}}}, \bibinfo
  {author} {\bibfnamefont {S.~P.}\ \bibnamefont {Neumann}}, \bibinfo {author}
  {\bibfnamefont {B.}~\bibnamefont {Liu}}, \bibinfo {author} {\bibfnamefont
  {T.}~\bibnamefont {Scheidl}}, \bibinfo {author} {\bibfnamefont {G.~C.}\
  \bibnamefont {Lorenzo}}, \bibinfo {author} {\bibfnamefont {{\v
  Z}.}~\bibnamefont {Samec}}, \bibinfo {author} {\bibfnamefont
  {L.}~\bibnamefont {Kling}}, \bibinfo {author} {\bibfnamefont
  {A.}~\bibnamefont {Qiu}}, \bibinfo {author} {\bibfnamefont {M.}~\bibnamefont
  {Razavi}}, \bibinfo {author} {\bibfnamefont {M.}~\bibnamefont {Stip{\v
  c}evi{\'c}}}, \bibinfo {author} {\bibfnamefont {J.~G.}\ \bibnamefont
  {Rarity}}, \ and\ \bibinfo {author} {\bibfnamefont {R.}~\bibnamefont
  {Ursin}},\ }\href {\doibase 10.1126/sciadv.aba0959} {\bibfield  {journal}
  {\bibinfo  {journal} {Sci. Adv.}\ }\textbf {\bibinfo {volume} {6}},\ \bibinfo
  {pages} {eaba0959} (\bibinfo {year} {2020})}\BibitemShut
  {NoStop}%
\bibitem [{\citenamefont {Fujii}\ \emph {et~al.}(2007)\citenamefont {Fujii},
  \citenamefont {Namekata}, \citenamefont {Motoya}, \citenamefont {Kurimura},\
  and\ \citenamefont {Inoue}}]{Fujii2007}%
  \BibitemOpen
  \bibfield  {author} {\bibinfo {author} {\bibfnamefont {G.}~\bibnamefont
  {Fujii}}, \bibinfo {author} {\bibfnamefont {N.}~\bibnamefont {Namekata}},
  \bibinfo {author} {\bibfnamefont {M.}~\bibnamefont {Motoya}}, \bibinfo
  {author} {\bibfnamefont {S.}~\bibnamefont {Kurimura}}, \ and\ \bibinfo
  {author} {\bibfnamefont {S.}~\bibnamefont {Inoue}},\ }\href {\doibase
  10.1364/OE.15.012769} {\bibfield  {journal} {\bibinfo  {journal} {Opt.
  Express}\ }\textbf {\bibinfo {volume} {15}},\ \bibinfo {pages} {12769}
  (\bibinfo {year} {2007})}\BibitemShut {NoStop}%
\bibitem [{\citenamefont {Zhu}\ \emph {et~al.}(2013)\citenamefont {Zhu},
  \citenamefont {Tang}, \citenamefont {Qian}, \citenamefont {Helt},
  \citenamefont {Liscidini}, \citenamefont {Sipe}, \citenamefont {Corbari},
  \citenamefont {Canagasabey}, \citenamefont {Ibsen},\ and\ \citenamefont
  {Kazansky}}]{Zhu2013}%
  \BibitemOpen
  \bibfield  {author} {\bibinfo {author} {\bibfnamefont {E.~Y.}\ \bibnamefont
  {Zhu}}, \bibinfo {author} {\bibfnamefont {Z.}~\bibnamefont {Tang}}, \bibinfo
  {author} {\bibfnamefont {L.}~\bibnamefont {Qian}}, \bibinfo {author}
  {\bibfnamefont {L.~G.}\ \bibnamefont {Helt}}, \bibinfo {author}
  {\bibfnamefont {M.}~\bibnamefont {Liscidini}}, \bibinfo {author}
  {\bibfnamefont {J.~E.}\ \bibnamefont {Sipe}}, \bibinfo {author}
  {\bibfnamefont {C.}~\bibnamefont {Corbari}}, \bibinfo {author} {\bibfnamefont
  {A.}~\bibnamefont {Canagasabey}}, \bibinfo {author} {\bibfnamefont
  {M.}~\bibnamefont {Ibsen}}, \ and\ \bibinfo {author} {\bibfnamefont {P.~G.}\
  \bibnamefont {Kazansky}},\ }\href {\doibase 10.1364/ol.38.004397} {\bibfield
  {journal} {\bibinfo  {journal} {Opt. Lett.}\ }\textbf {\bibinfo {volume}
  {38}},\ \bibinfo {pages} {4397} (\bibinfo {year} {2013})}\BibitemShut
  {NoStop}%
\bibitem [{\citenamefont {Peters}\ \emph {et~al.}(2003)\citenamefont {Peters},
  \citenamefont {Altepeter}, \citenamefont {Jeffrey}, \citenamefont
  {Branning},\ and\ \citenamefont {Kwiat}}]{Peters2005}%
  \BibitemOpen
  \bibfield  {author} {\bibinfo {author} {\bibfnamefont {N.}~\bibnamefont
  {Peters}}, \bibinfo {author} {\bibfnamefont {J.}~\bibnamefont {Altepeter}},
  \bibinfo {author} {\bibfnamefont {E.}~\bibnamefont {Jeffrey}}, \bibinfo
  {author} {\bibfnamefont {D.}~\bibnamefont {Branning}}, \ and\ \bibinfo
  {author} {\bibfnamefont {P.}~\bibnamefont {Kwiat}},\ }\href
  {http://portal.acm.org/citation.cfm?id=2011568} {\bibfield  {journal}
  {\bibinfo  {journal} {Quantum Inf. Comput.}\ }\textbf {\bibinfo {volume}
  {3}},\ \bibinfo {pages} {503} (\bibinfo {year} {2003})}\BibitemShut {NoStop}%
\bibitem [{\citenamefont {Poppe}\ \emph {et~al.}(2004)\citenamefont {Poppe},
  \citenamefont {Fedrizzi}, \citenamefont {Ursin}, \citenamefont {B\"{o}hm},
  \citenamefont {Lor\"{u}nser}, \citenamefont {Maurhardt}, \citenamefont
  {Peev}, \citenamefont {Suda}, \citenamefont {Kurtsiefer}, \citenamefont
  {Weinfurter}, \citenamefont {Jennewein},\ and\ \citenamefont
  {Zeilinger}}]{Poppe2004}%
  \BibitemOpen
  \bibfield  {author} {\bibinfo {author} {\bibfnamefont {A.}~\bibnamefont
  {Poppe}}, \bibinfo {author} {\bibfnamefont {A.}~\bibnamefont {Fedrizzi}},
  \bibinfo {author} {\bibfnamefont {R.}~\bibnamefont {Ursin}}, \bibinfo
  {author} {\bibfnamefont {H.~R.}\ \bibnamefont {B\"{o}hm}}, \bibinfo {author}
  {\bibfnamefont {T.}~\bibnamefont {Lor\"{u}nser}}, \bibinfo {author}
  {\bibfnamefont {O.}~\bibnamefont {Maurhardt}}, \bibinfo {author}
  {\bibfnamefont {M.}~\bibnamefont {Peev}}, \bibinfo {author} {\bibfnamefont
  {M.}~\bibnamefont {Suda}}, \bibinfo {author} {\bibfnamefont {C.}~\bibnamefont
  {Kurtsiefer}}, \bibinfo {author} {\bibfnamefont {H.}~\bibnamefont
  {Weinfurter}}, \bibinfo {author} {\bibfnamefont {T.}~\bibnamefont
  {Jennewein}}, \ and\ \bibinfo {author} {\bibfnamefont {A.}~\bibnamefont
  {Zeilinger}},\ }\href {\doibase 10.1364/OPEX.12.003865} {\bibfield  {journal}
  {\bibinfo  {journal} {Opt. Express}\ }\textbf {\bibinfo {volume} {12}},\
  \bibinfo {pages} {3865} (\bibinfo {year} {2004})}\BibitemShut {NoStop}%
\bibitem [{\citenamefont {Treiber}\ \emph {et~al.}(2009)\citenamefont
  {Treiber}, \citenamefont {Poppe}, \citenamefont {Hentschel}, \citenamefont
  {Ferrini}, \citenamefont {Lorünser}, \citenamefont {Querasser},
  \citenamefont {Matyus}, \citenamefont {H\"{u}bel},\ and\ \citenamefont
  {Zeilinger}}]{Treiber2009}%
  \BibitemOpen
  \bibfield  {author} {\bibinfo {author} {\bibfnamefont {A.}~\bibnamefont
  {Treiber}}, \bibinfo {author} {\bibfnamefont {A.}~\bibnamefont {Poppe}},
  \bibinfo {author} {\bibfnamefont {M.}~\bibnamefont {Hentschel}}, \bibinfo
  {author} {\bibfnamefont {D.}~\bibnamefont {Ferrini}}, \bibinfo {author}
  {\bibfnamefont {T.}~\bibnamefont {Lorünser}}, \bibinfo {author}
  {\bibfnamefont {E.}~\bibnamefont {Querasser}}, \bibinfo {author}
  {\bibfnamefont {T.}~\bibnamefont {Matyus}}, \bibinfo {author} {\bibfnamefont
  {H.}~\bibnamefont {H\"{u}bel}}, \ and\ \bibinfo {author} {\bibfnamefont
  {A.}~\bibnamefont {Zeilinger}},\ }\href@noop {} {\bibfield  {journal}
  {\bibinfo  {journal} {New J. Phys.}\ }\textbf {\bibinfo {volume} {11}},\
  \bibinfo {pages} {045013} (\bibinfo {year} {2009})}\BibitemShut {NoStop}%
\bibitem [{\citenamefont {Blume-Kohout}(2010)}]{Blume2010}%
  \BibitemOpen
  \bibfield  {author} {\bibinfo {author} {\bibfnamefont {R.}~\bibnamefont
  {Blume-Kohout}},\ }\href {http://stacks.iop.org/1367-2630/12/i=4/a=043034}
  {\bibfield  {journal} {\bibinfo  {journal} {New J. Phys.}\ }\textbf {\bibinfo
  {volume} {12}},\ \bibinfo {pages} {043034} (\bibinfo {year}
  {2010})}\BibitemShut {NoStop}%
\bibitem [{\citenamefont {Williams}\ and\ \citenamefont
  {Lougovski}(2017)}]{Williams2017}%
  \BibitemOpen
  \bibfield  {author} {\bibinfo {author} {\bibfnamefont {B.~P.}\ \bibnamefont
  {Williams}}\ and\ \bibinfo {author} {\bibfnamefont {P.}~\bibnamefont
  {Lougovski}},\ }\href {http://stacks.iop.org/1367-2630/19/i=4/a=043003}
  {\bibfield  {journal} {\bibinfo  {journal} {New J. Phys.}\ }\textbf {\bibinfo
  {volume} {19}},\ \bibinfo {pages} {043003} (\bibinfo {year}
  {2017})}\BibitemShut {NoStop}%
\bibitem [{\citenamefont {Lukens}\ \emph {et~al.}(2020)\citenamefont {Lukens},
  \citenamefont {Law}, \citenamefont {Jasra},\ and\ \citenamefont
  {Lougovski}}]{Lukens2020}%
  \BibitemOpen
  \bibfield  {author} {\bibinfo {author} {\bibfnamefont {J.~ M.}~\bibnamefont
  {Lukens}}, \bibinfo {author} {\bibfnamefont {K.~J.~H.}~\bibnamefont {Law}},
  \bibinfo {author} {\bibfnamefont {A.}~\bibnamefont {Jasra}}, \ and\ \bibinfo
  {author} {\bibfnamefont {P.}~\bibnamefont {Lougovski}},\ }\href@noop {}
  {\bibfield  {journal} {\bibinfo  {journal} {New J. Phys.}\ }\textbf {\bibinfo
  {volume} {22}},\ \bibinfo {pages} {063038} (\bibinfo {year}
  {2020})}\BibitemShut {NoStop}%
\bibitem [{\citenamefont {Vergyris}\ \emph {et~al.}(2017)\citenamefont
  {Vergyris}, \citenamefont {Kaiser}, \citenamefont {Gouzien}, \citenamefont
  {Sauder}, \citenamefont {Lunghi},\ and\ \citenamefont
  {Tanzilli}}]{Vergyris2017}%
  \BibitemOpen
  \bibfield  {author} {\bibinfo {author} {\bibfnamefont {P.}~\bibnamefont
  {Vergyris}}, \bibinfo {author} {\bibfnamefont {F.}~\bibnamefont {Kaiser}},
  \bibinfo {author} {\bibfnamefont {E.}~\bibnamefont {Gouzien}}, \bibinfo
  {author} {\bibfnamefont {G.}~\bibnamefont {Sauder}}, \bibinfo {author}
  {\bibfnamefont {T.}~\bibnamefont {Lunghi}}, \ and\ \bibinfo {author}
  {\bibfnamefont {S.}~\bibnamefont {Tanzilli}},\ }\href@noop {} {\bibfield
  {journal} {\bibinfo  {journal} {Quantum Sci. Technol.}\ }\textbf
  {\bibinfo {volume} {2}},\ \bibinfo {pages} {024007} (\bibinfo {year}
  {2017})}\BibitemShut {NoStop}%
\bibitem [{\citenamefont {Yoshizawa}\ and\ \citenamefont
  {Tsuchida}(2004)}]{Yoshizawa2004}%
  \BibitemOpen
  \bibfield  {author} {\bibinfo {author} {\bibfnamefont {A.}~\bibnamefont
  {Yoshizawa}}\ and\ \bibinfo {author} {\bibfnamefont {H.}~\bibnamefont
  {Tsuchida}},\ }\href@noop {} {\bibfield  {journal} {\bibinfo  {journal}
  {Appl. Phys. Lett.}\ }\textbf {\bibinfo {volume} {85}},\ \bibinfo {pages}
  {2457} (\bibinfo {year} {2004})}\BibitemShut {NoStop}%
\bibitem [{\citenamefont {{Finisar}}(2020)}]{FinisarWSS}%
  \BibitemOpen
  \bibfield  {author} {\bibinfo {author} {\bibnamefont {{Finisar}}},\
  }\href@noop {} {\enquote {\bibinfo {title} {Single wavelength selective
  switch ({WSS})},}\ }\bibinfo {howpublished}
  {\url{https://finisarwss.com/wp-content/uploads/2020/07/FinisarWSS_Single_Wavelength_Selective_Switch_ProductBrief_Jun2020.pdf}}
  (\bibinfo {year} {2020})\BibitemShut {NoStop}%
\bibitem [{\citenamefont {Lukens}\ and\ \citenamefont
  {Lougovski}(2017)}]{Lukens2017}%
  \BibitemOpen
  \bibfield  {author} {\bibinfo {author} {\bibfnamefont {J.~M.}\ \bibnamefont
  {Lukens}}\ and\ \bibinfo {author} {\bibfnamefont {P.}~\bibnamefont
  {Lougovski}},\ }\href {\doibase 10.1364/OPTICA.4.000008} {\bibfield
  {journal} {\bibinfo  {journal} {Optica}\ }\textbf {\bibinfo {volume} {4}},\
  \bibinfo {pages} {8} (\bibinfo {year} {2017})}\BibitemShut {NoStop}%
\bibitem [{\citenamefont {{Lu}}\ \emph {et~al.}(2019)\citenamefont {{Lu}},
  \citenamefont {{Weiner}}, \citenamefont {{Lougovski}},\ and\ \citenamefont
  {{Lukens}}}]{IEEEptl2019}%
  \BibitemOpen
  \bibfield  {author} {\bibinfo {author} {\bibfnamefont {H.-H.}\ \bibnamefont
  {{Lu}}}, \bibinfo {author} {\bibfnamefont {A.~M.}\ \bibnamefont {{Weiner}}},
  \bibinfo {author} {\bibfnamefont {P.}~\bibnamefont {{Lougovski}}}, \ and\
  \bibinfo {author} {\bibfnamefont {J.~M.}\ \bibnamefont {{Lukens}}},\
  }\href@noop {} {\bibfield  {journal} {\bibinfo  {journal} {IEEE Photon.
  Technol. Lett.}\ }\textbf {\bibinfo {volume} {31}},\ \bibinfo {pages} {1858}
  (\bibinfo {year} {2019})}\BibitemShut {NoStop}%
\bibitem [{\citenamefont {Sandoval}\ \emph {et~al.}(2019)\citenamefont
  {Sandoval}, \citenamefont {Lingaraju}, \citenamefont {Imany}, \citenamefont
  {Leaird}, \citenamefont {Brodsky},\ and\ \citenamefont
  {Weiner}}]{Sandoval2019}%
  \BibitemOpen
  \bibfield  {author} {\bibinfo {author} {\bibfnamefont {O.~E.}\ \bibnamefont
  {Sandoval}}, \bibinfo {author} {\bibfnamefont {N.~B.}\ \bibnamefont
  {Lingaraju}}, \bibinfo {author} {\bibfnamefont {P.}~\bibnamefont {Imany}},
  \bibinfo {author} {\bibfnamefont {D.~E.}\ \bibnamefont {Leaird}}, \bibinfo
  {author} {\bibfnamefont {M.}~\bibnamefont {Brodsky}}, \ and\ \bibinfo
  {author} {\bibfnamefont {A.~M.}\ \bibnamefont {Weiner}},\ }\href@noop {}
  {\bibfield  {journal} {\bibinfo  {journal} {Opt. Lett.}\ }\textbf {\bibinfo
  {volume} {44}},\ \bibinfo {pages} {1674} (\bibinfo {year}
  {2019})}\BibitemShut {NoStop}%
\bibitem [{\citenamefont {Lingaraju}\ \emph {et~al.}(2020)\citenamefont
  {Lingaraju}, \citenamefont {O’Malley}, \citenamefont {Jones}, \citenamefont
  {Sandoval}, \citenamefont {Azzouz}, \citenamefont {Leaird}, \citenamefont
  {Lukens}, \citenamefont {Brodsky},\ and\ \citenamefont
  {Weiner}}]{Lingaraju2020}%
  \BibitemOpen
  \bibfield  {author} {\bibinfo {author} {\bibfnamefont {N.~B.}\ \bibnamefont
  {Lingaraju}}, \bibinfo {author} {\bibfnamefont {N.}~\bibnamefont
  {O’Malley}}, \bibinfo {author} {\bibfnamefont {D.~E.}\ \bibnamefont
  {Jones}}, \bibinfo {author} {\bibfnamefont {O.~E.}\ \bibnamefont {Sandoval}},
  \bibinfo {author} {\bibfnamefont {H.~N.}\ \bibnamefont {Azzouz}}, \bibinfo
  {author} {\bibfnamefont {D.~E.}\ \bibnamefont {Leaird}}, \bibinfo {author}
  {\bibfnamefont {J.~M.}\ \bibnamefont {Lukens}}, \bibinfo {author}
  {\bibfnamefont {M.}~\bibnamefont {Brodsky}}, \ and\ \bibinfo {author}
  {\bibfnamefont {A.~M.}\ \bibnamefont {Weiner}},\ }in\ \href@noop {} {\emph
  {\bibinfo {booktitle} {CLEO: QELS}}}\ (\bibinfo {organization} {Optical
  Society of America},\ \bibinfo {year} {2020})\ p.\ \bibinfo {pages}
  {FF1D.5}\BibitemShut {NoStop}%
\end{thebibliography}

\end{document}